\documentstyle[12pt,aps,prc,preprint,epsfig]{revtex}
\tightenlines

\begin{document}

\newpage
\title {
 Discovery of Long-Lived Shape Isomeric States which Decay by Strongly
Retarded High-Energy Particle Radioactivity
}
\author{ A. Marinov$^{(1)}$, S. Gelberg$^{(1)}$, and D. Kolb$^{(2)}$}
\address{$^{(1)}$ Racah Institute of Physics, The Hebrew University, Jerusalem
91904, Israel}
\address{$^{(2)}$ Department of Physics, Kassel University, 34109 Kassel,
Germany}
\maketitle
\begin{abstract}
The reaction $^{28}$Si + $^{181}$Ta has been studied at E$_{Lab}$=125
and 135 MeV. Coincidences between
high energy particles and various
X- and $\gamma$-rays from abnormally long-lived states were observed.
e.g.
7.8-8.6 MeV $\alpha$-particles with $\gamma$-rays of a superdeformed band,
5.1-5.5 MeV $\alpha$-particles with X- and $\gamma$-rays of W, Re, and Pt,
and 3.88 MeV particles
(interpreted as protons) with 185.8 keV $\gamma$-rays. The
data are interpreted in terms of the production of long-lived (t$_{1/2}$
of several months) high spin isomeric states in the
 second well of the potential in the parent nuclei, which decay to the normal
 states in the daughters, and in the third well of the potential, which
 decay to the second well.

\end{abstract}


\section{Introduction}

The first evidence for a new kind of long-lived isomeric states was
obtained in actinide fractions  \cite{1,2}
produced via secondary reactions
in a CERN W target which had been irradiated with 24-GeV protons
 \cite{3}.  Isomeric states with t$_{1/2}$ $\sim$ 0.6 y and  $\geq$  30 d
(10$^{4}$ - 10$^{5}$ times longer than the
 expected half-lives of the corresponding ground states) were found in
neutron-deficient
 $^{236}$Am and $^{236}$Bk nuclei, respectively.  About 3x10$^{5}$
atoms of $^{236}$Am and 4x10$^{4}$ atoms of $^{236}$Bk were produced
in the isomeric states, and decayed by the $\beta^{+}$ or electron capture
processes. The character of these states was not clear:
they are far from closed shells where high spin isomers are usually found, and
 they have very long lifetimes as compared to the known shape isomers.

In addition \cite{2,4,5}, some new
particle groups were
found in the decay of various actinide fractions separated from the same W
target.
For example, an unambiguous 5.14 MeV $\alpha$-particle group
with t$_{1/2}$ $\sim$ 4 y  was seen in the Bk source \cite{1,2,4,5,6}.
This half-life is a factor of 2x10$^{6}$ -- 3x10$^{3}$ too short
 \cite{7} for the normal
$\alpha$-decay of Bk -- Pu nuclei. It was very difficult to understand such an
enhancement. Furthermore, 3.0 and 4.0 MeV particle groups were seen from the
Am source in coincidence with L$_{\alpha1}$ X-rays in the Am region \cite{2}.
 Since the  relationship between the particle energies and their lifetimes
deviate by about 23 and 12 orders \cite{4} from the systematics \cite{7}
of $\alpha$-particles, it was assumed \cite{4} that they were protons of
unknown origin.

The clue to the understanding of the above mentioned findings has been
obtained recently in
several studies \cite{8,9} of the $^{16}$O + $^{197}$Au reaction at
E$_{Lab}$ = 80 MeV, where similar isomeric states and particle decays
were found.  An isomeric state which decays by emitting a
 5.20 MeV $\alpha$-particle with t$_{1/2}$ $\sim$ 90 m has been found in
 $^{210}$Fr.   Since this half-life is longer than the known
half-life of the ground state of $^{210}$Fr, it was concluded that a long-lived
 isomeric state had been formed in this nucleus. A t$_{1/2}$ of 90 m
for 5.20 MeV $\alpha$-particles in $^{210}$Fr is enhanced by a factor
 of 3x10$^{5}$ as compared to normal transitions \cite{7,8}.  However, this
group was observed in coincidence with $\gamma$-rays which fit
predictions for a super-deformed band \cite{8}.  Therefore the effect of large
deformations of the nucleus on the $\alpha$-particle decay was
calculated and found \cite{8} to be consistent with the observed enhancement.
It was argued \cite{8} that since the isomeric state decays to a high spin
state,
it should also have high spin, and since it decays by
 enhanced $\alpha$-particle emission
to state(s) in the second well of the potential, it should be in the second
 well itself.

The predicted \cite{10,11} excitation energies of the second minima
in the evaporation residue nuclei and their daughters
produced by the $^{16}$O+$^{197}$Au reaction are above the proton separation
energies. Therefore a search for long-lived proton decays has been
performed \cite{9} using the same reaction.  Two long-lived proton activities
 with half-lives of about 6 h and 70 h were found \cite{9} with proton
energies of 1.5 -- 4.8 MeV with a  sharp line at 2.19 MeV.
This energy may correspond to a predicted transition \cite{10}
from the second minimum in $^{198}$Tl to the ground state of $^{197}$Hg
with E$_p$ = 2.15 MeV. The superdeformed state in $^{198}$Tl may be produced
in an alpha decay chain of transitions from shape isomeric states to
shape isomeric states starting from $^{210}$Fr. It should be noted  that
a long-lived high-spin isomeric state in the second minimum of the
potential in $^{241}$Pu has been predicted by
S. G. Nilsson et al. \cite{12} back in 1969.

The present work reports on the discovery of new long-lived isomeric states,
 produced
by the $^{28}$Si + $^{181}$Ta reaction at bombarding energies of 125
and 135 MeV, which decay by strongly hindered $\alpha$-particle and proton
transitions.
The lower energy is about $10\%$ below the Coulomb barrier.
A fusion cross section of about 10 mb is predicted for this energy using a
coupled-channel deformation code \cite{13} with deformation parameters
$\beta_2$=0.41 for $^{28}$Si and $\beta_2$=0.26 for $^{181}$Ta
 \cite{14},\footnote{The $\beta_2$ value
for $^{181}$Ta was taken as the average from the corresponding values of
$^{180}$Hf and $^{182}$W.}
and allowing for 2$^+$ and 3$^-$ excitations in $^{28}$Si. Only
2 $\mu$b is predicted when no deformations are included in the calculations.
For 135 MeV the corresponding
predicted fusion cross sections are 95 mb with deformations
and 40 mb without. At bombarding energies of 125 and 135 MeV, the compound
nucleus is formed at excitation energies of 42.1 and 50.8 MeV, respectively.
Preliminary results of this work have been published before
\cite{15,6,16}.

\section{Experimental Procedure}

In three irradiations, 2.5 mg/cm$^{2}$ Ta targets followed by stacks of C catcher
foils were bombarded with a $^{28}$Si beam obtained from the Pelletron
accelerator in Rehovot.  Irradiations I and II were
performed with a 125 MeV
$^{28}$Si beam and irradiation III with 135 MeV.
Carbon catcher foils
of 200 $\mu$g/cm$^{2}$ and 60 $\mu$g/cm$^{2}$ were used in order to
catch the evaporation residue nuclei and their daughters.
In all experiments the average beam intensity
was about 11 pnA and the total dose about 1x10$^{16}$ particles.  Long
period off-line measurements were carried out in the laboratory in Jerusalem,
using the irradiated catcher foils as sources.
In the present paper we summarize
the results of the particle-$\gamma$ coincidence measurements.\footnote
{$\alpha$-$\alpha$ correlation measurements were performed with
60 $\mu$g/cm$^{2}$
 C catcher foils situated in between two 300 $\mu$g/cm$^{2}$ Si detectors,
but no $\alpha$ correlation events with $\Delta$t$\leq$ 10 s could be
 significantly established. The preliminarily claimed
correlations \cite{15}
cannot presently be ruled out to be due to other physical or electronic
background
effects.  At longer correlation times the number of random events was too
large.}
  A 450 mm$^{2}$, 300 $\mu$m thick, Si
surface barrier detector and a 500 mm$^{2}$, 10 mm thick, thin
 window Ge(Li) detector were used for these measurements.
The Si detector was calibrated using the 3.18 MeV $\alpha$-particle
group
of a $^{148}$Gd source and an accurate pulse generator.
The Ge(Li) detector was calibrated using
the X- and $\gamma$-rays of a $^{57}$Co source.
The source was sandwiched between the Si and the Ge detector.
 Two 0.01 inch thick Be foils separated the source from the Ge detector.
 The first foil
 was the window for the vacuum chamber which
included the source and the Si detector, and the second one the window of
the Ge detector. The transparency of the two Be windows together was close to
100\% for gamma-rays with E$_\gamma$ $\geq$ 10 keV.
 The solid angle
of the Si detector was about $34\%$ of 4$\pi$ sr and of the Ge(Li)
detector $16\%$.
The peak-to-total ratio of the Ge(Li) detector was 100\% up to about
120 keV and decreased gradually to 22\% at 250 keV. Its
FWHM energy resolution was about  900 eV.
The intrinsic resolution of the Si detector was about 25 keV.
The full line widths for $\alpha$-particles of 5.0 to 8.6 MeV
were 0.52 to 0.34 MeV,
respectively, with a 200 $\mu$g/cm$^{2}$ C foil, and
 the corresponding values for the 60 $\mu$g/cm$^{2}$ foil
were 0.16 to 0.10 MeV.
The resolving time of the coincidence system was 1 $\mu$s. The coincidence
events together with the singles particle events were recorded event by event
with a time accuracy of 1 ms.

\section{Results}
\subsection{Singles and Coincident Events}
Figs.~1 and 2 show typical singles particle and $\gamma$-ray spectra obtained
 at 125 MeV bombarding energy.  Figs.~3 and 4 show similar spectra obtained at
135 MeV beam energy.  From the measured $\alpha$-particle and $\gamma$-ray
 energies and lifetimes, production cross sections for various evaporation
residue nuclei, with estimated errors of $\pm$25\%, were deduced and are
summarized in table 1. (Cross sections for
some isomeric states (see below), are also given in table 1.)

Figs.~5 and 6 give respective $\alpha$-$\gamma$ coincidence plots from
measurements I and II, corresponding to irradiations I and II.
Figs.~7 and 8 are one dimensional projection plots on the $\alpha$-particle
axis of figs.~5 and 6, respectively, for E$\gamma$ $\geq$ 25 keV.
Fig.~9 presents a particle-$\gamma$ coincidence spectrum
from measurement III, corresponding to irradiation III.\footnote{During
the
first 77 d after irradiation I was completed, a search for proton activity
using the
$\Delta$E-E system of Ref. \cite{9} was performed and gave negative results with
 an upper limit
 of about 0.5 nb for half-lives of between 20 hours to 70 days.
 On the other hand, in the 135 MeV experiment
(fig. 9) we most probably saw protons in
 the particle-$\gamma$ coincidences. (See further below).}
The estimated numbers of random coincidences in the 3 -- 10 MeV particle energy
and 10 -- 250 keV $\gamma$-energy
are 5.4 x 10$^{-2}$, 4.4 x 10$^{-2}$, and
0.4 in figs. 5, 6, and 9, respectively.
Some of the coincidence events in figs. 5, 6 and 9, between
5-6 MeV $\alpha$-particles and various $\gamma$-rays,
may perhaps be due to a
contamination from an emanated $^{212}$Pb source from $^{228}$Th, which was
used in the same chamber about
4 months earlier \cite{9}.
Despite the long time delay
to our present long term measurements of 77 to 235 days,
and careful cleaning of the chamber, we might have picked up some residual
contamination.
Also, indication for K X-rays of Rn is seen in this region. While in principle
their origin may be from some reaction products, they also may be from the
decay of $^{223}$Ra, which belongs to the $^{235}$U chain. Although it is
not likely, but because of this ambiguity, we will not try to make a claim
about this region. A two-dimensional background measurement,
taken for 8 days,
before the $^{212}$Pb source was used in the chamber, gave zero events in
the whole region of 3.5 -- 10 MeV particles and 0 -- 250 keV $\gamma$'s.

It is seen in figs.~5 and 6, and figs.~7 and 8, that while the region of
$\alpha$-particles of 6 -- 8 MeV is relatively empty, coincidence events
are seen between
 various $\gamma$-rays and $\alpha$-particles of 8 -- 9 MeV.
In fig.~5,  13 coincidence events between
 7.99 - 8.61 MeV $\alpha$-particles and various $\gamma$-rays of E $\geq$
 20 keV are
 observed, where 8 of them fall within a narrow range of 190 keV particle
energy,
 between 8.42 - 8.61 MeV.  (The estimated $\alpha$-particle full line width
for the 200 $\mu$g/cm$^2$ C foil is around
340 keV). From the intensities of the singles in the $\alpha$- and $\gamma$-ray
spectra (339 $\alpha$-particles and 2.14x10$^{7}$ gamma-rays in 76.8 days),
and the resolving time of 1~$\mu$s of the coincidence system,
the total number of random coincidences in the 8 - 9 MeV
 region was estimated to be 2.2x10$^{-3}$.  In fig. 6  one finds 8 coincidence
events between 8.19 -- 9.01 MeV, where 5 of them fall within
70 keV, between 8.55 -- 8.62 MeV, and one at 9.01 MeV.  (The $\alpha$-particle
full line
width for a 60 $\mu$g/cm$^2$ C foil is about 100 keV).
The corresponding estimated total number of random coincidences between
8 to 9 MeV is
1.4x10$^{-3}$. One does not see such concentration of events in the same energy
region in fig. 9 which was taken for about 2.7 times longer period
than figs. 5 and 6.
3 events are seen in the 8 - 9 MeV region in fig. 9 which sets an upper
limit on the background in figs. 5 and 6 to be aroud 1 count as compared
to the respective 13 and 8 counts seen experimentally.
It was estimated that in the 8 -- 9 MeV range
of $\alpha$-particles,
 at most 0.8 events in fig.~5 and 0.3
events in fig.~6 may be due to the contamination mentioned
above of
coincidences between
$\beta^{-}$ and $\gamma$-rays from $^{212}$Bi and the 8.78 MeV
 $\alpha$-particles from $^{212}$Po (t$_{1/2}$ = 0.3 $\mu$s).

Fig.~10 shows time sequence plots for the $\alpha$-$\gamma$ coincidences with
$\alpha$-energies of 8--9 MeV obtained in measurements I and II and seen in
 figs.~5 and 6, respectively.
In measurement II a growth in the intensity of the 8--9 MeV group was
found from the beginning
up to 97 days after the end of irradiation. In measurement I no significant
change in decay rate was observed between 77 to 154 days and, by binning the data,
t$_{1/2}$ $\geq$ 40 days
 was estimated for the lower limit of the decaying half-life.
An additional measurement, taken about 4.5 y after the one presented in
fig. 5 and measurement I of fig. 10, gave 0 counts in 22.0 d. An upper
limit for the half-life of t$_{1/2}$ $\leq$ 2.1 y is deduced from this
measurement.
 \subsection{Rotational Bands and Sum Events}
Table 2(a) and fig.~11 show that almost
all the coincidence events in figs. 5 and 6 with 7.8-8.61 MeV
 $\alpha$-particles (the encircled ones) fit very nicely with a J(J+1) law
assuming
E$_{x}$ = 4.42xJ(J+1) keV and $\Delta$J = 1.
 This fit is very
significant from a statistical point of view \cite{8}. All the gamma-rays in
fig.~5
(7 out of 7 events), and almost all of them in fig. 6 (6 out of 7 events),
which are above 90 keV, fit, to within $\pm$0.5 keV on the average, with this
formula. The probability for 13 out of 14 events which are distributed
evenly (for instance due to Compton effect from higher energy gamma-rays),
to fall into 13 specified energy positions, within 1 keV is:

$\mbox{$(^{14}_{13})$p$^{13}$(1-p)$^{(14-13)}$ = 4.1 x 10$^{-8}$}$.

p is equal to 36x1.0/(250-90). 36 is the number of possible
gamma-transitions between states
of both integer and half integer spins with E$_{x}$ = 4.42 x J(J+1),
 in the range of 90 to 250 keV.
(The four low energy events of E$_{\gamma}$ $\leq$ 60 keV in fig. 5
and the one event in fig. 6 which do not fit with the J(J+1) rule may be
due to Compton events).
It is also seen in fig. 6 and in table 2(b) that the $\gamma$-rays of the
3 coincidence events
 in the region of 6 - 7 MeV $\alpha$-energy of 6.16 MeV - 176.1 keV,
 6.94 MeV - 207.4 keV
 and 6.41 MeV - 242.3 keV,
 fit with 20 $\rightarrow$ 19 (176.4), 47/2 $\rightarrow$ 45/2 (207.3)
and 55/2 $\rightarrow$ 53/2 (242.6) J $\rightarrow$ J-1 transitions, assuming
E$_{x}$ = 4.41xJ(J+1)
 keV.  The probability
in this case that this fit is accidental due to a chance coincidence of
evenly distributed events is 1.2x10$^{-3}$, taking into account that p =
18x0.47/(250-170).
18 is the number of possible gamma-transitions in the range of 170 to 250
keV, and 0.47 keV is twice the average deviation from the specified energies.
(This probability increases to 1.1x10$^{-2}$ if, instead of 0.47 keV, the
value which was used before of 1.0 keV is assumed in the calculation).
 It should be mentioned that rotational constants
 around 4.4 keV
correspond to superdeformed
bands [SDB] in this region of nuclei \cite{8}.

In fig. 9 one sees about 10 scattered events in the 6.5 -- 9 MeV region. Some
of them may also fit with a J(J+1) law. However, the statistical significance
of the fit in this case is not so good and we will not discuss them further.

The appearance in table 2a
of both integer and half-integer spins at low $\gamma$-ray energies
indicates that bands
in both even
and odd nuclei were formed.\footnote{It should be mentioned that identical
bands in neighboring nuclei are known \cite{17}.}
 (At high energies the observed transitions are
probably sum events. See the following paragraphs.)
$^{207}$Rn, produced via pn evaporation, and
its daughters, are, from the kinematic point of view, the best
 candidates for the half-integer transitions.  The
 8.42 MeV $\alpha$-particle in coincidence with a photon of 67.1 keV (seen
in fig. 5 and table 3 below), which fits with K$_{\alpha1}$ X-ray of
 Pt (known energy 66.832 keV), may indicate that the $\alpha$-transition
is from
$^{195}$Hg to $^{191}$Pt.  $^{195}$Hg is the daughter after 3 $\alpha$-particle
decays from $^{207}$Rn.

It should be mentioned that the high energy $\gamma$-rays (above about 120 keV,
table 2)
are most probably sum events rather than photo-peak events. If, for instance,
 the three coincidence events in fig. 6 at E$_\gamma$ of 247.0 (2 events)
 and 225.3 keV are photo-peak events, then
 9.5$\pm$5.5 Compton events below 120 keV (the position of the Compton edge)
should have been seen
 in coincidence with $\alpha$-particles of
8.55-8.61 MeV, but only one event is seen.
Similarly if the $\gamma$-rays which are in coincidence with 6-7 MeV
$\alpha$-particles, seen in the same figure and mentioned above, are
photo-peak events,
then 7.6$\pm4.4$
Compton events should have been seen in this $\alpha$-energy range at
E$_{\gamma}$ $\leq$ 120 keV, while only one event is seen.
On the other hand, for a
 de-excitation of a superdeformed band with $\Delta$J = 1, a measured energy
which fits a transition between high-spins, may be due to many sum combinations
 of lower energy transitions between lower spins, for which low-energy Compton
events should not be observed.  For instance, an energy of 225.4 keV which
 fits a 51/2 $\rightarrow$ 49/2 transition (assuming E$_{x}$ = 4.42$\times$J
(J+1) keV, table 2) may be due to one of 17 different
combinations of 3-fold events, if only the first 14 levels of the band up to
maximum spin of 27/2, and maximum transition energy of 119.3 keV are considered,
for which the photo-peak to total ratio is 100\%.
(119.3 (27/2 $\rightarrow$ 25/2) + 92.8 (21/2 $\rightarrow$ 19/2) + 13.3
(3/2 $\rightarrow$ 1/2) = 225.4, or 84.0 (19/2 $\rightarrow$ 17/2) + 75.1 (17/2
$\rightarrow$ 15/2) + 66.3 (15/2 $\rightarrow$ 13/2) = 225.4 are
two examples out of
these 17 possible combinations). Because of the
large solid angle of the $\gamma$-detector (0.16) and the large number
of possible 3-fold combinations, the probability P$_{3}$
to see
a
3-fold event is larger than the probability P$_{1}$
to see a 1-fold event.
(These probabilities
for the above example are:
P$_{1}$ = $\epsilon$'(1-$\epsilon$)$^{(N-1)}$ =
0.042(1-0.16)$^{24}$ = 6.4 x 10$^{-4}$,
and P$_{3}$ = n$_{3}$$\epsilon$$^{3}$(1-$\epsilon$)$^{N-3}$~=~
17x0.16$^{3}$(1-0.16)$^{10}$ =
1.2x10$^{-2}$, about 20 times larger. $\epsilon$' is the photo-peak
efficiency of the detector for the particular energy,
$\epsilon$ is the
geometrical efficiency, N is the number of transitions in the band counting
from the level to which the $\alpha$-particle
is decaying, and n$_{3}$ is the number of 3-fold combinations).

Many combinations of 2- or 4- fold events of a half-integer spin band
will give energies which fit with those of a band of an integer spin. For
instance, the two events at 247.0 keV (table 2) which fit in energy to a
28 $\rightarrow$ 27 transition, and the 186.5 keV event which fits with a
21 $\rightarrow$ 20 transition,  may be due to such combinations, since the
$\alpha$-particle energy is the same as that of the half-integer transitions
of 225.3 and 171.8
keV, and therefore they presumably belong to the same band. (In  an
integer-spin band of $\Delta$J = 1 various combinations of 2-, 3-
and
4-fold events
of low energy transitions have large efficiencies and give energies which fit
with those of transitions between higher spins. This property, both for
half-integer and integer spin bands, does not exist for $\Delta$J = 2
transitions).

As shown above the detection of sum events which are energetically degenerate
with the diagram line leads to a substantially increased effective detection
efficiency (dependent, however, on the maximum available spin) for this energy.
The minimum spin which is consistent with the data
is 19/2. The largest efficiency of the detector (which corresponds to the
smallest production cross section), taking into account 2-
and 4-fold sum events of transitions between half-integer spins in order to
get the 247.0 keV transition, is obtained with a spin of 35/2.
The lower limit for the production cross section
estimated from Irradiation II and taking into account the detector efficiency
for various possible combinations
of sum events, as described above, is about 130 nb
(table 1).\footnote{Previously \cite{15,16}, it was assumed that
the 247.0 (2 events),
225.3 (2 events) and 191.7 keV events, which are in coincidence with 8-9 MeV
$\alpha$-particles, are photo-peak events due to known transitions
in $^{195}$Hg and $^{197}$Tl.
However, in this case
12.6$\pm$6.3 Compton events
below 120 keV should have been seen in Fig. 6 in coincidence
with $\alpha$-particles
of 8.55-8.62 MeV, but only one event is seen.}
\footnote{In principle one
may think that the observed
$\gamma$-energies are Compton events due to higher energy $\gamma$-rays.
But then the probability \cite{8} to have two events in the same
position (within the
energy resolution) is only a few percent, and this occurred twice, at 247.0
(fig. 6) and 225.3 keV (figs. 5 and 6).}

  As shown above, the observation in
table 2 of gamma rays which fit in energies with transitions between high
spin SDB states is consistent with various sums of gamma rays due to
transitions between lower spin states with $\Delta$J = 1.  Since the
 intensity of the characteristic X-rays seen in figs. 5 and 6 in coincidence
with 8-9 and 6-7 MeV
$\alpha$-particles is very low,
the observed
$\gamma$-rays should be due to E1 transitions, with quite low conversion
 factors,
rather than M1 ones.  Such transitions have been seen before \cite{8} and were
predicted for SD wells in nuclei around Z = 86 and N = 116 \cite{18,19,20}.
\subsection{Coincidences between $\alpha$-Particles and Characteristic
X-Rays and Identified $\gamma$-Rays}
In order to identify some of the coincidence events seen in figs. 5, 6, and
9 at lower particle energies below 6 MeV, we first looked on groups of events.
The existence of groups of 3 or 2 coincidence events in clean regions
indicates that they are probably not background. Secondly, one looks on
their accurately measured photon energies. At low energies the observed photon
may be either an X-ray or a $\gamma$-ray. The energies of the characteristic
X-rays of the elements are known very well \cite{23}. The K$_{\alpha1}$ X-rays
of the various elements of the evaporation residue nuclei and their daughters
are separated one from another by about 2 keV, where the value of a
K$_{\alpha1}$-line of an element with Z protons is about the same as the value
of the K$_{\alpha2}$-line of an element with Z+1 protons. If the measured
energy turns out to be very close to a known X-ray energy, then it is
reasonable to
assume that the observed photon is an X-ray rather then a $\gamma$-ray. (In
the cases found by us and discussed below the deviations from
the known values are from 0.03 to 0.28 keV, while the width of one channel
in the $\gamma$-spectrum is about 0.24 keV and the FWHM is about
four channels). The X-ray may be emitted if the $\alpha$-particle is decaying
to an excited state which  decays by conversion electrons. Such a process
is followed by emitting characteristic X-rays of the daughter nucleus. The
Z-value of the daughter is thus determined. In order to determine its
A-value we tried to identify the $\gamma$-rays which are in coincidence
with $\alpha$-particles of about the same energy. A consistency is obtained
when the $\gamma$-ray fit in energy with the measured value, belongs to
an isotope of the determined element, and it is consistent with known
or expected values of the conversion factors, and also with the decay scheme.

In several cases, as seen below, some of the observed events
have been identified as due to characteristic X-rays (the events surrounded
with squares in figs.~5, 6 and 9) and as known $\gamma$-transitions in various
nuclei (the events
surrounded with triangles in figs.~5, 6 and 9).  In fig.~6 two coincidences between
 about 5.50 MeV $\alpha$-particles and 61.1 and 59.4 keV photons are seen,
and fit with K$_{\alpha1}$ and K$_{\alpha2}$ X-rays of Re of 61.140
 and 59.718 keV.  At about the same $\alpha$-energy two 144.0 keV events
are observed and can be identified with a  known transition in $^{186}$Re.  This
 identification is supported by the observation in fig.~5, in the same
$\alpha$-energy range, of coincidence events with respective 59.7 keV and
140 keV X- and
 $\gamma$-rays. The first
fits with  K$_{\alpha2}$ of Re, and the second with a transition in
 $^{186}$Re, which precedes the above 144.0 keV transition.
A coincidence between 5.45 MeV $\alpha$-particles and 59.4 or 59.7 keV photons
may in principle be also due to the known decay of $^{241}$Am of 5.486 MeV
$\alpha$ and 59.54 keV $\gamma$. However, we did not use and never had in
our laboratory a $^{241}$Am calibration source. Under these circumstances
the observation of the 61.1 keV photon, which fits very nicely
with K$_{\alpha1}$ of Re, and which is very far off (1.56 keV) from the
$\gamma$-rays of $^{241}$Am, and of the two consecutive $\gamma$-rays
of $^{186}$Re, suggests that the two events at about 59.6 keV are due to
K$_{\alpha2}$ of Re and not due to $^{241}$Am. A 59.6 keV X-ray may in
principle be also due to K$_{\alpha1}$ of W of 59.3182 keV. Here also
the observation  of the other X-rays and $\gamma$-rays suggest that
it is K$_{\alpha2}$ of Re rather than K$_{\alpha1}$ of W. Out of total
number of 3 X-rays, one is expecting to see 1.9 K$_{\alpha1}$ events and
1.1 K$_{\alpha2}$ events. The observation of 1 event in the first case and 2
in the second, is well within the statistical error.

In fig. 9,
for $\alpha$-particles between 5.18 and 5.53 MeV, groups of two and three
events with photon energies of 66.8 keV and 59.6 keV, respectively, are seen.
They may correspond to the known K$_{\alpha1}$ X-rays of Pt of 66.832 keV
and of W of 59.318 keV of W.
  At the same corresponding
$\alpha$-energies known $\gamma$-rays of $^{189}$Pt and $^{183}$W are
observed. Here the three 59.6 keV events are more likely to be due to
K$_{\alpha1}$ of W, rather than K$_{\alpha2}$ of Re. If they were
K$_{\alpha2}$ of Re then 5.2 events of K$_{\alpha1}$ at 61.140 keV
are expected. The observation of zero events when 5 are expected is unlikely.
On the other hand, if the 3 observed events are K$_{\alpha1}$ X-rays of W, then
1.7 events of K$_{\alpha2}$ of W at 57.98 keV are expected.
The observation of zero events when 1.7 are expected is well within the
statistical errors.

As mentioned above and seen in table 3, about the same $\alpha$-energy
corresponds to two different decays. One leads to $^{186}$Re and was obtained
in the E$_{Lab}$ = 125 MeV experiment, and the other presumably leads to
$^{183}$W and was obtained in the E$_{Lab}$ = 135 MeV experiment. As seen below
(section IVA), these data are interpreted as due to transitions from the second
well of the potential in the parent nuclei to the normal states in the
daughters. There is nothing against having about the same $\alpha$-particle
energy in two different transitions.
Because of the low statistics we consider the identifications mentioned above
as tentative only, where the identification of $^{186}$Re is better than of the
other.

  The results are summarized in table 3 together with possible reaction
channels and corresponding decay chains which lead to the observed
transitions. The
deduced production cross sections (with an accuracy of about a factor
of 2) are given in table 1, columns 3 and 5. In
general lower limits were deduced since the branching ratios along the
decay chains are not known.

 Some
 coincidence events
between 6.17 -- 9.01 MeV $\alpha$-particles and identified X-rays with no
 corresponding $\gamma$-rays were seen.  They are also given in table 3.
\subsection{Evidence for Proton Radioactivity}
A very well defined coincidence group of three events is seen in fig.~9
at an average particle energy of 3.88 MeV and $\gamma$-energy of 185.8 keV.
Because of the low energy of the particles and the very narrow total width
of 40 keV, one may conclude that the particles are protons.  (The total
 estimated widths at this energy are 55 keV for protons
  and 630 keV for $\alpha$-particles.  The
estimated half-life for 3.88 MeV $\alpha$-particles decaying to, for instance,
$^{194}$Hg is about 1x10$^{8}$ y \cite{7}). The 185.8 keV $\gamma$-rays fit
with a
known transition in $^{204}$Rn (table 3). The production cross section of
this group is given in table 1.

\section{Discussion}
\subsection{Transitions from Superdeformed to Normal States}
The E$_{\alpha}$ values for the g.s. to g.s. transitions for $^{190}$Ir
$\rightarrow$ $^{186}$Re;  $^{187}$Os $\rightarrow$ $^{183}$W and
 $^{193}$Hg $\rightarrow$ $^{189}$Pt mentioned in table 3 are 2.74, 2.662 and
2.927 MeV \cite{22}, respectively.  The corresponding observed
 $\alpha$-energies of 5.43, 5.53 and
 5.18 MeV (table 3) are clearly due to decay of isomeric states in the parent
 nuclei.  The estimated half-lives for normal $\alpha$-particles of these
 energies \cite{7} are 2.9x10$^{3}$;  3.5x10$^{2}$ and 1.2x10$^{6}$ s.  The
 observed lifetimes of several months, which may be due to combined lifetimes
along the long decay chains, are retarded by 3 and 4 orders of magnitude
 in the first and second case, respectively.

From the above mentioned measured $\alpha$-energies the lower limits for the
excitation energies of the isomeric states in the parent nuclei $^{190}$Ir,
 $^{187}$Os and
$^{193}$Hg (see table 3) are deduced to be 3.1; 3.0 and 4.5 MeV, respectively.
(Lower limits are deduced since the $\alpha$-decay may proceed through a higher
excitation energy than the one seen experimentally).
Extrapolated and interpolated predicted energies \cite{10,11} for the second
minima  of 4.1 \cite{10} or 4.2 \cite{11}; 3.4 \cite{10} or 3.6 \cite{11}; and
 4.2 \cite{10} or 4.6 \cite{11} MeV, respectively, are also given in table 3.
The observed energies of the isomeric states and the
predicted positions of the second minima seem to be in the same range of
excitation energies.  This suggests that the isomeric states
in these cases are in the second well of the potential, and that the $\alpha$
transitions are from the superdeformed well in the parent nuclei to less
 deformed or normal states in the daughters.  These are the last
transitions after several decays, presumably from isomeric state to isomeric
 state, all within the second wells along the decay chain.  For instance,
$^{190}$Ir in the isomeric state may be produced after 6 decays, 4$\alpha$
and 2$\beta$$^{+}$ or EC decays, all from superdeformed to
superdeformed states (see table 3).\footnote{More complicated situations
where the evaporation residue nucleus is produced in the third (hyperdeformed)
well of the potential, and then decays to the second well by $\alpha$-particle
or even by proton emission, is in principle not impossible.}
 Since the last step of the chain is a
decay to a relatively high spin state, it seems that the originally produced
isomeric state in the evaporation residue nucleus has high spin, and that the
transitions from mothers to daughters along the decay chain are between high
spin states.

\subsection{Transitions from Superdeformed to Superdeformed States?}
We now discuss the three coincidence events seen in fig.~6 and mentioned
above (Secion III.B. and table 2b) of 6.16~MeV -- 176.1~keV;
6.94~MeV -- 207.4~keV
and  6.41 -- 242.3~keV.
These events appeared 96, 68 and 78 days respectively, after the end of
irradiation, and their lifetimes therefore are of the order of several months.
  As shown above, the energies of the first two $\gamma$-rays fit
very nicely with SDB transitions between half-integer spins, where
the most probable candidates are
 $^{207}$Rn, produced by the pn evaporation process, and its daughters.  The
coincidence event of a 6.17~MeV $\alpha$-particle with 79.1 keV photon (fig.~5
 and table 3) which fits with Po K$_{\alpha1}$ X-rays of 79.290 keV, suggests
that the transitions mentioned above with 6 -- 7 MeV $\alpha$-particles
might be
 from isomeric state(s) in $^{207}$Rn to the SDB states in the second well of
 the potential in $^{203}$Po. (The half-life of the ground state of
$^{207}$Rn is 9.3~m).  The measured $\alpha$-particle energies are in the
predicted range for the transition from the second minimum in $^{207}$Rn
to the second minimum in $^{203}$Po of 6.2 \cite{10} or 6.9 \cite{11} MeV.

Similarly the energies of  7.16 MeV (fig. 6; E$_{Lab}$ =
125 MeV) and 7.05 MeV (fig. 9; E$_{Lab}$ = 135 MeV) $\alpha$-particles in
coincidence  with L$_{\beta1}$(At) (see table 3) may correspond to
 the predicted $\alpha$-particle transitions between the second minima of
 $^{206}$Fr$^{s.m.}$ and $^{202}$At$^{s.m.}$ (3n reaction) of 6.9
\cite{10} or 7.3 \cite{11} MeV, and of $^{205}$Fr$^{s.m.}$
to $^{201}$At$^{s.m.}$
 (4n reaction) of 7.0 \cite{10} and 7.2 \cite{11} MeV, respectively.

In both cases discussed in the present subsection (IV.B.),
while the $\alpha$-energies are consistent with this picture, the very long
 lifetime may indicate that the internal structure of the parent isomeric
states and
the final rotational SDB states are different.
\subsection{Proton Transition from a Superdeformed to a Normal State?}
For the 3.88 proton group (Section III.D. above), if the identification
given in table 3 is correct,
the excitation energy of the isomeric state in $^{205}$Fr is $\geq$ 6.7 MeV
(see table 3).  This is quite high as compared to 3.9 \cite{10} or 5.0
 \cite{11} MeV predicted
 for the second minimum, and may indicate that the origin of the isomeric state
 in this case is different.
\subsection{Hyperdeformed to Superdeformed Transitions}
The most striking result is the $\alpha$-particle group around 8.6 MeV which was
 found in
coincidence with SDB transitions (figs. 5, 6, 7, 8, and table 2a).
As mentioned
above the appearance of $\gamma$-transitions between half-integer spins
shows that most of these transitions are in an odd A nucleus, where
$^{207}$Rn and its daughters are most probable
candidates from the reaction kinematic point of view.\footnote{As shown in
Section III.B. above, the high energy integer spin transitions of 247.0 and
186.5 keV (table 2a) are presumably due to sum combinations of half-integer
transitions as well.}  The observation of
 a coincidence
event between a 8.42 MeV $\alpha$-particle (which is, within the experimental
spread for $\alpha$-particles, consistent with 8.6 MeV for 200 $\mu$g/cm$^{2}$
C catcher foil) and a K$_{\alpha1}$ X-ray of Pt (fig. 5 and table 3) indicates
that the transition may be from $^{195}$Hg (produced after 3 $\alpha$-decays
from $^{207}$Rn) to $^{191}$Pt. However, transitions from other Hg isotopes
produced
by fewer $\alpha$-decays, which are followed by several $\beta$$^{+}$ or EC
decays cannot be excluded.

The predicted \cite{7} half-life for normal 8.6 MeV $\alpha$-particles is below
1 $\mu$s. Therefore in principle the $\gamma$-rays of the SDB transitions
may either precede or follow
the 8.6 MeV $\alpha$-particles.\footnote{In all the cases
discussed above the first possibility
is excluded since the lifetime of the $\alpha$-particles is always much
longer than 1$\mu$s, and they could not be detected in coincidence
with $\gamma$-rays which precede them.}
In the case where the $\gamma$-rays come first, the $\alpha$-particles are due
to either a transition from the second minimum in the parent nucleus to the
second minimum in the daughter, or from the second minimum in the parent to
 normal states in the daughter.  However in the first case the energy
 of 8.6 MeV is much larger than the predicted 3.3 MeV in
$^{195}$Hg$^{s.m.}$ $\rightarrow$ $^{191}$Pt$^{s.m.}$ transition \cite{10,11},
 and 6.3 \cite{10} or 6.9\cite{11} in $^{207}$Rn$^{s.m.}$ $\rightarrow$
 $^{203}$Po$^{s.m.}$ transition.   On the other hand in the second case of
an $\alpha$-transition from the second minimum to normal states,
the $\alpha$-particle lifetime will most probably be retarded, and the 8.6 MeV
$\alpha$-particles will not be in coincidence within 1 $\mu$s with the SDB
 transitions.
(Furthermore, the predicted transition energy from the second minimum to the
ground state in
 $^{195}$Hg$^{s.m.}$ $\rightarrow$ $^{191}$Pt$^{g.s.}$ is about
7.2 MeV \cite{10,11}
 and in $^{207}$Rn$^{s.m.}$ $\rightarrow$ $^{203}$Po$^{g.s.}$ is 6.3
 \cite{10} or 6.9 MeV \cite{11}.  These values are also not in accord with
the experimental value of 8.6 MeV).

It is therefore reasonable to assume that the 8.6 MeV
$\alpha$-particles decay first and then are followed by the SDB $\gamma$-rays,
  similar to the situation
in all the other cases discussed above.
 Let us first consider the case where the isomeric state is in the second
 well of the parent nucleus which decays by the 8.6 MeV $\alpha$-particles
to the SDB in the daughter.  One faces here two problems, namely the
very large hindrance factor of the $\alpha$-particles, and their high energy.
The hindrance factor of the $\alpha$-particles is in the range of
 10$^{16}$ (t$_{1/2}$ $\sim$ 40 d as compared to the predicted value of
 1.4 x 10$^{-10}$ s assuming $\beta$$_{2}$ = 0.7 \cite{8}).
As for the energy, if, for instance, the 8.6 MeV $\alpha$-particles decay to
 a
SDB state of spin 27/2 in $^{191}$Pt at E$_{x}$ =  865~keV
(E$_{0}$ = 4.42 keV), and taking
into account the
predicted excitation energy of the second minimum in $^{191}$Pt at about
4.1 MeV \cite{10,11}, and the Q$_\alpha$(g.s.$\rightarrow$g.s.) value for
$^{195}$Hg of 2.190 MeV \cite{22}, the isomeric state in $^{195}$Hg
turns out to be
 at about 11.6 MeV above the g.s.
Assuming an  excitation energy of the second minimum in $^{195}$Hg
 of  about 5.2 MeV \cite{10,11}, the above energy of 11.6 MeV corresponds to
an
 excitation energy of about 6.4 MeV above the second minimum, which seems
unlikely. (In the same way, about 5.0 and 3.0 MeV excitation energies above
the second minimum
are respectively deduced
for more neutron-rich Hg isotopes and for $^{207}$Rn, which is unlikely
as well.)

  It seems reasonable to conclude that the large hindrance factor
  might be due to another barrier transition,
  as, for instance,
from the third well (the hyper-deformed well)
\cite{24,25,26} to the second one.\footnote{It should also be mentioned that
a superdeformed minimum
on the oblate side, which might decay to the prolate superdeformed minimum
via triaxial
shapes, has been predicted in
$^{236}$Cm at $\beta_{2}$ = --0.63 and E$_{x}$ = 8.2 MeV, using Hartree-Fock
calculations
\cite{27}.}  It should be mentioned that an extrapolated value of about 11.5 MeV
is obtained for the excitation energy of the third minimum in $^{195}$Hg
from the predictions of Ref. \cite{25}.  This value is in accord with the
  deduced value mentioned above of about 11.6 MeV.\footnote{The retardation of
the $\alpha$-events of 6-7 MeV discussed in section IV.B. above is unlikely
to be due to transitions from the third minimum, as its predicted excitation
energy of about
18.8 MeV (extrapolated from Ref.\cite{26}) comes out far from the data of 7.8
or 10.3 MeV (table 3).}     The large
 production cross
 section of the isomeric state which decays by the 8.6 MeV $\alpha$-particles
of about 130 nb, as compared to the production of the isomeric states in the
second minima of a few nb (see table 1), may also indicate that its origin is
from the third minimum, which can presumably be produced more easily at
bombarding
energies below the Coulomb barrier.

It is also seen in table 1 that while the highest
 cross section for
the production of the normal evaporation residue nuclei is about a factor of
60 larger at 135 MeV as compared to 125 MeV, the production of the isomeric
states in the second minima is about the same at these two bombarding energies,
 whereas there is
an
 exceptionally large cross section at
125 MeV for the 8.6 MeV $\alpha$-particle group which is in coincidence with
the SDB transitions.

\section{Summary}
In summary, evidence for long-lived isomeric states with lifetimes of up to
several
months, with abnormal decay properties,
 was obtained.  Relatively high energy $\alpha$-particles of 5.1-5.5
MeV were tentatively found in the usually non-$\alpha$-emitting nuclei, $^{187}$Os,
$^{190}$Ir and $^{193}$Hg, and interpreted as due to transitions from the
second well of the potential in the parent nuclei to less deformed or normal
 states in the daughters.
The observed transitions are the last in long chains of up to 6 steps
of $\alpha$-particle and $\beta^{+}$ or EC transitions, presumably within the
second well itself.

Long-lived 6 -- 7 MeV $\alpha$-particles,
in coincidence with SDB $\gamma$-rays, were found with energies consistent
with transitions from the second minimum in the parent
nucleus to the second minimum in the daughter. The reason for their long
lifetimes though is not clear.

 A 3.88 MeV long-lived proton group was found. Its state of
origin is not entirely clear.

By far the most exciting  observation is the very
high energy and strongly retarded (t$_{1/2}$ $\geq$ 40 d) 8.6 MeV
$\alpha$-particle group in
coincidence with SDB transitions, which is interpreted as due to a long-lived
isomeric state in $^{195}$Hg at E$_{x}$ $\cong$ 11.6 MeV, and is consistent
with a transition from
  the third (hyperdeformed) well of the parent nucleus to
 the second (superdeformed) well in the daughter.

The production cross
sections of the isomeric states in the second well were about the same
 at E$_{Lab}$ = 125 (10\% below the barrier) and at E$_{Lab}$ = 135 MeV, while
on the average a factor of 15 -- 60 larger cross section was found for the
 8.6 MeV group at 125 MeV.

The existence of such long-lived isomeric states, with lifetimes much longer
than those of their corresponding ground states, and with preferential
production
cross sections at relatively low bombarding energies, may add new considerations
regarding the stability and the production mechanism of heavy and superheavy
elements \cite{3,28,29,30}. In particular it should be mentioned that the
extra-push energies needed for the  production of such nuclei in their
superdeformed and hyperdeformed wells are much smaller than expected for
producing them in their normal states \cite{31}.
\section{Acknowledgements}
We acknowledge the support of the accelerator crew of the Weizmann
Institute at Rehovot, and the technical assistance of S. Gorni, O.
Skala and the electronic team of the Racah Institute. D.K.
acknowledges the financial support of the DFG.  We are grateful to
N.~Zeldes and J. L. Weil for very valuable discussions.

\newpage
\begin{table}[h]
\caption {Production cross sections for various evaporation
residue nuclei obtained in their ground states and in the isomeric
states via the $^{28}$Si + $^{181}$Ta reaction at 125 and 135
MeV.}
\begin{center}
\begin{tabular}{lllll}
Reaction   & $\sigma$$^{g.s.}$($\mu$b) & \mbox{\boldmath
$\sigma$$^{i.s.}$}{\bf(nb)} &  $\sigma$$^{g.s.}$($\mu$b)  &
\mbox{\boldmath $\sigma$$^{i.s.}$}{\bf(nb)} \\
     & 125 MeV & {\bf 125 MeV}   & 135 MeV
     & {\bf135 MeV}  \\ \hline \\
  3n      & $\leq{3.4}$     & \mbox{\boldmath $\geq{8}$$^{a}$}
& $\leq{155}$$^{b}$  &
  \\
p2n     & 4.5             & & $\leq{23}$$^{b}$   &  \\ 4n      &
&        & 274  & \mbox{\boldmath $\leq{9}$$^{d}$}  \\ p3n     &
&        & 101  & \mbox{\boldmath $\geq{3}$$^{a}$}  \\ 5n      &
&        & $\leq{21}$$^{c}$   &   \\ p4n     &                 &
& $\leq{21}$$^{c}$ &  \\ $\alpha$2n &              &        &
& \mbox{\boldmath $\geq{2}$$^{a}$}  \\ pn      &                 &
\mbox{\boldmath $\geq{130}$$^{a}$} &        &   \\
\end{tabular}
\end{center}
\end{table}

\noindent$^{a}$ Lower limit is given since the branching ratios
along the decay chain (table 3) are not known.\\ $^{b}$ It was
impossible to distinguish between the 3n and the p2n reactions.
The value given was deduced assuming that the relevant observed
activity is due to this reaction channel only.\\ $^{c}$ It was
impossible to distinguish between the 5n and the p4n reactions.
The value given was deduced assuming that the relevant observed
activity is due to this reaction channel only.\\ $^{d}$ Based on
the coincidence group of the 3.88 MeV protons with 185.8 keV
$\gamma$-rays and assuming that the $\gamma$-rays are from
$^{204}$Rn (see text).\\

\newpage
\begin{table}[h]
\caption { The energies of the $\gamma$-rays in coincidence with
the 7.8~-~8.6 (a) and 6.0~-~7.0 MeV (b) $\alpha$-particles (the
encircled events in Figs.~5 and 6), as compared to calculated
transitions assuming E$_{x}$ = E$_{0}$~x~J(J+1).\\}
\begin{center}
\begin{tabular}{lllll}
E$_{\alpha}$   & E$_{\gamma}$(expt.)$^{a}$    & E$_{\gamma}$(cal.)
& $\triangle$E  & Transition$^{b}$ \\ (MeV)  & (keV)     & (keV)
& (keV) \\ \hline \\ \multicolumn{5}{c}{a)
E$_{\alpha}$~=~7.8~-~8.6 MeV; E$_{x}$~=~4.42~x~J(J+1) keV
 }\\
7.81        & 25.9         & 26.5        & --0.6     &
3$\rightarrow$2 \\ 8.52        & 40.1         & 39.8        &
+0.3     & 9/2$\rightarrow$7/2 \\ 8.21        & 52.9         &
53.0        & --0.1     & 6$\rightarrow$5 \\ 8.53$^{c}$  &
97.9$^{c}$   & 97.2        &  +0.7     & 11$\rightarrow$10 \\ 8.19
& 118.4        & 119.3       & --0.9     & 27/2$\rightarrow$25/2
\\ 7.88        & 123.4        & 123.8       & --0.4     &
14$\rightarrow$13 \\ 8.37        & 141.3        & 141.4       &
--0.1     & 16$\rightarrow$15 \\ 7.99        & 145.6        &
145.9       & --0.3     & 33/2$\rightarrow$31/2 \\ 8.51        &
171.8        & 172.4       & --0.6     & 39/2$\rightarrow$37/2 \\
8.60        & 186.5        & 185.6       &  +0.9     &
21$\rightarrow$20 \\ 8.07        & 212.9        & 212.2       &
+0.7     & 24$\rightarrow$23 \\ 8.59$^{d}$  & 225.3$^{d}$  & 225.4
& --0.1     & 51/2$\rightarrow$49/2 \\ 8.61$^{e}$  & 247.0$^{e}$
& 247.5       & --0.5     & 28$\rightarrow$27 \\
\multicolumn{5}{c}{b) E$_{\alpha}$~=~6.0~-~7.0 MeV;
E$_{x}$~=~4.41~x~J(J+1) keV
 }\\
6.16        & 176.1        & 176.4       &  -0.3     &
20$\rightarrow$19 \\ 6.94        & 207.4        & 207.3       &
+0.1     & 47/2$\rightarrow$45/2 \\ 6.41        & 242.3        &
242.6       &  -0.3     & 55/2$\rightarrow$53/2 \\

\end{tabular}

\end{center}

\end{table}

\noindent$^{a}$ The peak to total ratio was 100\% up to about
120~keV and
 reduced
gradually to 22\% at 250~keV. \\ $^{b}$ The transitions between
the maximum possible spins are given.
 At relatively high energies the observed $\gamma$-rays are most probably sum
events due to various combinations of lower energy transitions in
the band which, because of the J(J+1) law, fit in energy to
calculated transitions between higher energies. (see text).\\
$^{c}$ Average of 2 events:  8.47~MeV $\alpha$ in coincidence with
97.4~keV $\gamma$ in the first measurement and 8.58~MeV $\alpha$
in coincidence with 98.3~keV $\gamma$ in the second measurement.
\\ $^{d}$ Average of 2 events:  8.61~MeV $\alpha$ in coincidence
with 224.9~keV $\gamma$ in the first measurement and 8.57~MeV
$\alpha$ in coincidence with 225.7~keV $\gamma$ in the second
measurement. \\ $^{e}$ Average of 2 events in the second
measurement of 8.62~MeV $\alpha$ in coincidence with 246.5~keV
$\gamma$ and 8.60~MeV $\alpha$ in coincidence with 247.5~keV
$\gamma$. \\

\newpage
\begin{table}[h]
\caption{Summary of tentatively identified X- and $\gamma$-rays
which were in coincidence with various particles in measurements I
and II (125~MeV) and measurement III (135~MeV). In column 1 we
give the measured $\gamma$-energies,
  their times of occurrence
after the end of irradiation or estimated half-life if several
events are observed (bold letters), and the measurement number. In
column 2 the respective particle energies are given. Column 3 are
the tentatively identified nuclei with the character of transition
in column 4 [23]. The fifth column gives the possible reaction
channel and the decay chain (bold letters) to the identified
mother isotope. (More complicated situations are in principle not
impossible. See text, section IVB). The sixth column shows lower
limits on the excitation energies of the decaying isomeric states
in the mother nuclei (in brackets).
 Theoretical predictions for the
excitation energies of the second minima (S.M.) in the mother
nuclei given by Satula et al.
 (S, ref. [10])
and Krieger et al. (K, ref. [11]) are in the last column. (For the
last case a prediction [25] for the third minimum is given).}
\begin{center}
\begin{tabular}{lllllll}

E$_{\gamma}$(Meas. No.)  & E$_{\alpha,p}$ & Isotope & Transition &
Reaction & E$^{I.S.}_{x}$($^{\rm {A}}$Z) &E$^{S.M.}_{x}$
 \\
keV  & MeV &  & & {\bf Decay } & MeV  & MeV
 \\
{\bf Time/d}$^{a}$ & & & & {\bf Chain} & & S; K \\  \hline \\
61.1(II)        & 5.48    &            & K$_{\alpha1}$(Re)      &
&     & \\ {\bf 31.1}  &         &            &
&   &     & \\ 59.6(x2)(I; II)    & 5.45    &            &
K$_{\alpha2}$(Re)      &   &
   &  \\
{\bf  7.4;~107.2}   &            &                   &    &   &
&  \\ 144.0(x2)(II)   & 5.43    & $^{186}$Re  &
330(?)$^{b}$$\rightarrow$186 & 3n   & $\geq$3.1($^{190}$Ir) &
4.1$^{d}$; 4.2$^{d}$ \\ {\bf  60.2;  88.8} &     &            &
(5$^{+}$)$\rightarrow$(6)$^{-}$ &{\bf 4\mbox{\boldmath$\alpha$},}
&    &  \\
    &       &         &  E1; 100\% &
{\bf 2\mbox{\boldmath$\beta$$^{+}$}(EC)$^{c}$} &    & \\ 141.0(I)
& 5.29    & $^{186}$Re  & 471(?)$^{b}$$\rightarrow$330(?)$^{b}$ &
3n        & $\geq$3.2($^{190}$Ir) & 4.1$^{d}$; 4.2$^{d}$ \\ {\bf
130.2 }  &        &            & (4$^{+}$)$\rightarrow$(5$^{+}$)
&{\bf 4\mbox{\boldmath$\alpha$},}            &
&  \\
   &        &         & M1+E2       &
{\bf 2\mbox{\boldmath$\beta$$^{+}$}(EC)$^{c}$}   & & \\
 66.8(x2)(III)  & 5.18 &   & K$_{\alpha1}$(Pt) &  &  &    \\
{\bf 13.9;  58.6}  &  &  &  &  &  & \\ 236.2(III)  & 5.12 &
$^{189}$Pt  & 2291.8$\rightarrow$2056.1 & p3n; 4n & $\geq$4.53
($^{193}$Hg)  & 4.2$^{e}$; 4.6$^{e}$    \\ {\bf 139.5}  &   &
&($\frac{29}{2}$$^{+}$)$\rightarrow$($\frac{27}{2}$$^{-}$) &
\mbox{\boldmath $3\alpha; 3\alpha,$} & & \\
    &  &  &  (E1)
& {\bf ---~;} \mbox{\boldmath$\beta$$^{+}$}{\bf(EC)}$^{c}$ &  & \\
59.6(x3)(III) & 5.53  & & K$_{\alpha1}$(W) &  &  &    \\ {\bf
t$_{\bf{1/2}}$ = 90}  & & & & & & \\ 210.3(III)  &  5.43 &
$^{183}$W  & 308.9$\rightarrow$99.1 &  $\alpha$2n &
$\geq$3.04($^{187}$Os)  &  3.4$^{d}$; 3.6$^{d}$   \\ {\bf 168.4}
&  &  &  $\frac{9}{2}$$^{-}$$\rightarrow$$\frac{5}{2}$$^{-}$ &
\mbox{\boldmath\ $4\alpha,$} &  &  \\
 &  &  &  E2; 100\%       &  \mbox{\boldmath$\beta$$^{+}$}{\bf(EC)}$^{c}$
  &  &   \\
70.9(II) & 9.01 & & K$\alpha_1$(Hg) & & & \\ {\bf 89.7} & & & & &
& \\ 67.1(I) & 8.42 & & K$\alpha_1$(Pt) & & & \\ {\bf 81.7} & & &
& & & \\ 79.1(I) & 6.17 & & K$\alpha_1$(Po) & & & \\ {\bf 133.5} &
& & & & & \\ 13.9(II) & 7.16 & & L$\beta_1$(At) & & & \\ {\bf
64.6} & & & & & & \\ 13.9(III) & 7.05 & & L$\beta_1$(At) & & & \\
{\bf 127.8} & & & & & & \\ 185.8(x3)(III)  &  3.88(p)  &
$^{204}$Rn  & 2219.0$\rightarrow$2032.8$^{f}$ &
  4n &  $\geq$6.7($^{205}$Fr) &  3.9$^{e}$; 5.0$^{e}$  \\
{\bf t$_{\bf{1/2}}$ = 120}  &  &  &
(9$^{-}$)$\rightarrow$(8$^{+}$) & --  &   & \\
  &  &  &  E1; 55\%  &  &  &    \\
SDB(II)$^{g}$ & 6 - 7 & $^{203}$Po& SDB & pn &
7.8$^{h}$;10.3$^{i}$($^{207}$Rn)& 6.8$^{d}$; 9.9$^{e}$  \\ {\bf
t$_{\bf{1/2}}$ = 53} &  &  & table 2(b) & --- &  \\ SDB(I;
II)$^{j}$ & 8.6 & $^{191}$Pt$^{k}$ & SDB & pn
 & $\sim$11.6($^{195}$Hg)$^{k}$ & 11.5($^{III min}$)$^{l}$ \\
{\bf t\mbox{\boldmath$_{1/2} \geq$} 40} &  &  & table 2(a) & {\bf
3}\mbox{\boldmath$\alpha$}& &\\
\end{tabular}
\end{center}
\end{table}

\noindent$^{a}$ For a group of three events the half-lives as
estimated according to the formulas of K.-H.~Schmidt et al. [21]
are given. \\ $^{b}$ The excitation energy in $^{186}$Re is not
certain [23]. \\ $^{c}$ The order of the decay is not known. \\
$^{d}$ Extrapolated value. \\ $^{e}$ Interpolated value. \\ $^{f}$
The identification given in the table is not certain as explained
in the text. \\ $^{g}$ See text and table 2(b) for the
$\gamma$-eneries of this band. \\ $^{h}$ Estimated excitation
energy in $^{207}$Rn assuming E$_{\alpha}$=6.5 MeV, a predicted
E$_{x}$ of the second minimum in $^{203}$Po of 6.6 MeV [10], and
decay to a SDB state with a spin of 27/2 at excitation energy
above the second minimum of 863 keV.\\ $^{i}$ As comment (h) above
except that E$_{x}$ of the second minimum in $^{203}$Po was taken
at 9.1 MeV as predicted in Ref. [11].\\ $^{j}$ See table 2(a) for
the $\gamma$-energies of this band. \\ $^{k}$ See text.\\ $^{l}$
Extrapolated value for the excitation energy of the third minimum
from Ref. [25] is given.


\newpage
\begin{figure}
\begin{center}
\leavevmode \epsfysize=8.0cm \epsfbox{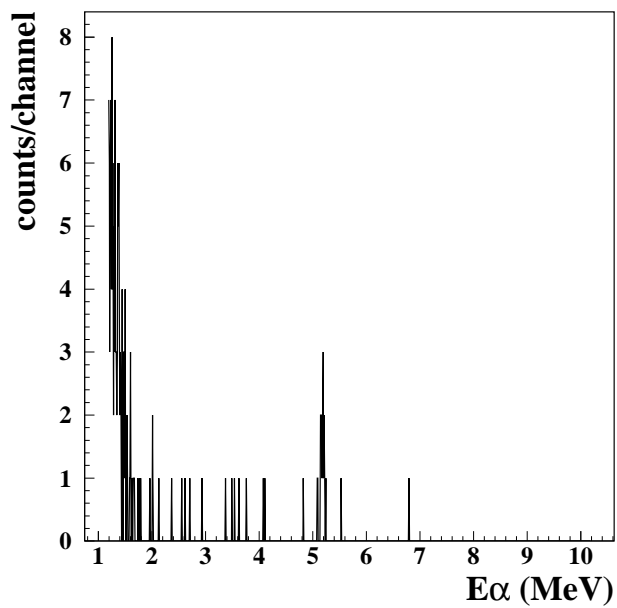}
\end{center}
\caption{Typical particle singles spectrum at
E$_{Lab}$($^{28}$Si)=125 MeV with 60 $\mu$g/cm$^2$ C catcher foil,
taken for 29 hours starting 3 hours after the end of irradiation.}
\end{figure}
\begin{figure}
\begin{center}
\leavevmode \epsfysize=8.0cm \epsfbox{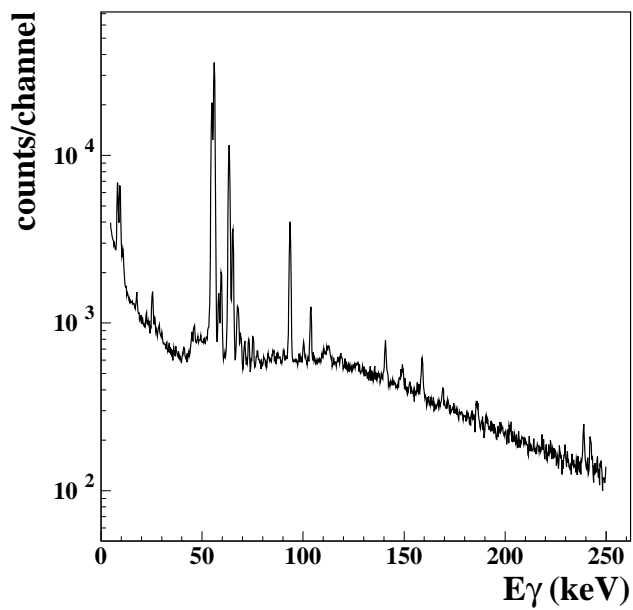}
\end{center}
\caption{Typical $\gamma$-ray singles spectrum. Same conditions as
fig. 1.}
\end{figure}
\begin{figure}
\begin{center}
\leavevmode \epsfysize=8.0cm \epsfbox{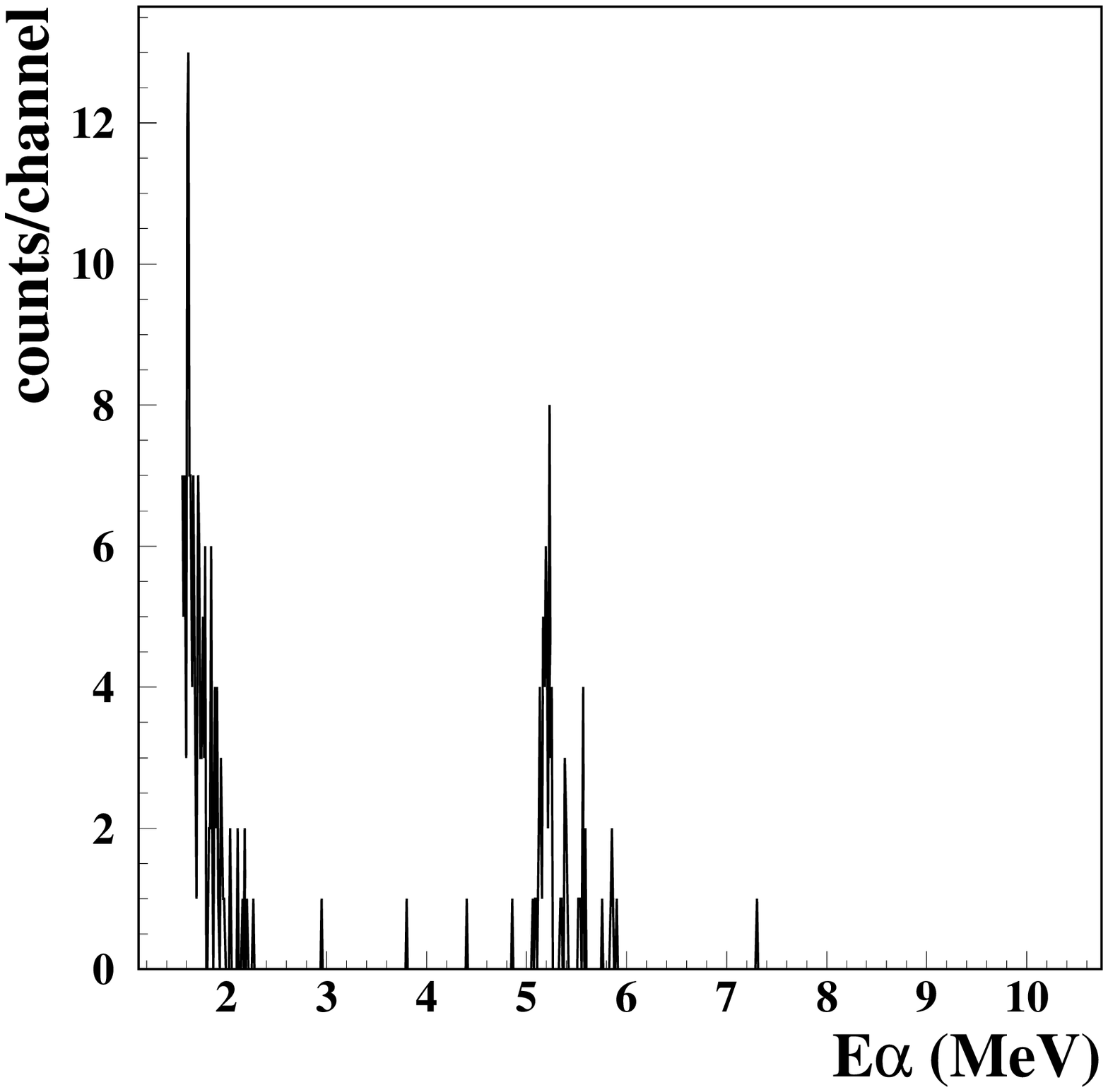}
\end{center}
\caption{Typical particle singles spectrum at
E$_{Lab}$($^{28}$Si)=135 MeV with 60 $\mu$g/cm$^2$ C catcher foil,
taken for 11 hours starting 3 hours after the end of irradiation.}
\end{figure}
\begin{figure}
\begin{center}
\leavevmode \epsfysize=8.0cm \epsfbox{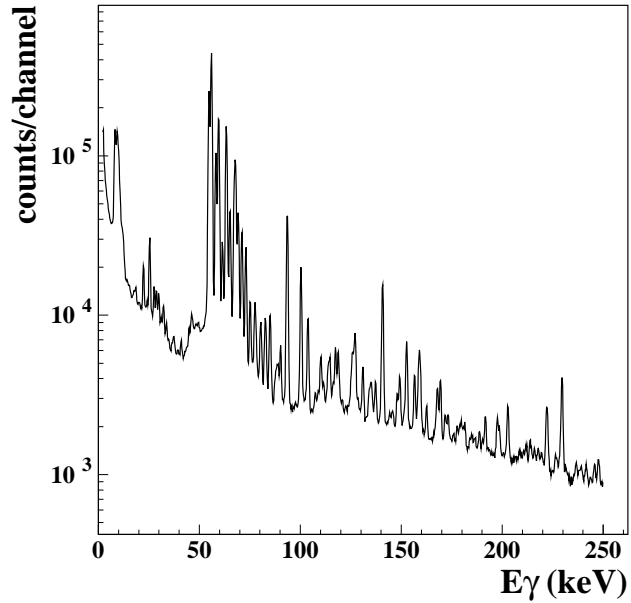}
\end{center}
\caption{Typical $\gamma$-ray singles spectrum at
E$_{Lab}$($^{28}$Si)=135 MeV with 200 $\mu$g/cm$^2$ C catcher
foil, taken for 24.5 hours, starting 3 hours after the end of
irradiation.}
\end{figure}
\begin{figure}
\begin{center}
\leavevmode \epsfysize=8.0cm \epsfbox{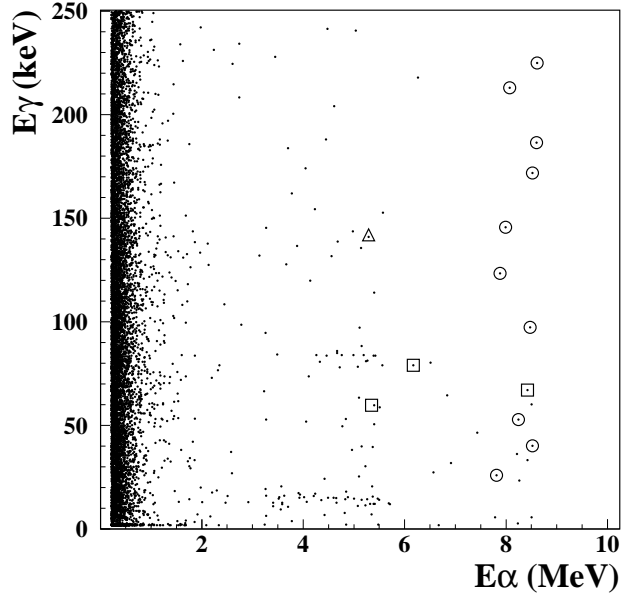}
\end{center}
\caption{Particle-$\gamma$ coincidences from measurement I:
E$_{Lab}$($^{28}$Si)=125 MeV, with 200 $\mu$g/cm$^2$ C catcher
foil, taken for 76.8 days, starting 77.4 days after the end of
irradiation. The $\gamma$-ray energies of the encircled events fit
with SDB transitions. The photon energies of the events in squares
fit with known characteristic X-rays. The $\gamma$-ray energies of
the events in triangles are identified with known transitions.}
\end{figure}
\begin{figure}
\begin{center}
\leavevmode \epsfysize=8.0cm \epsfbox{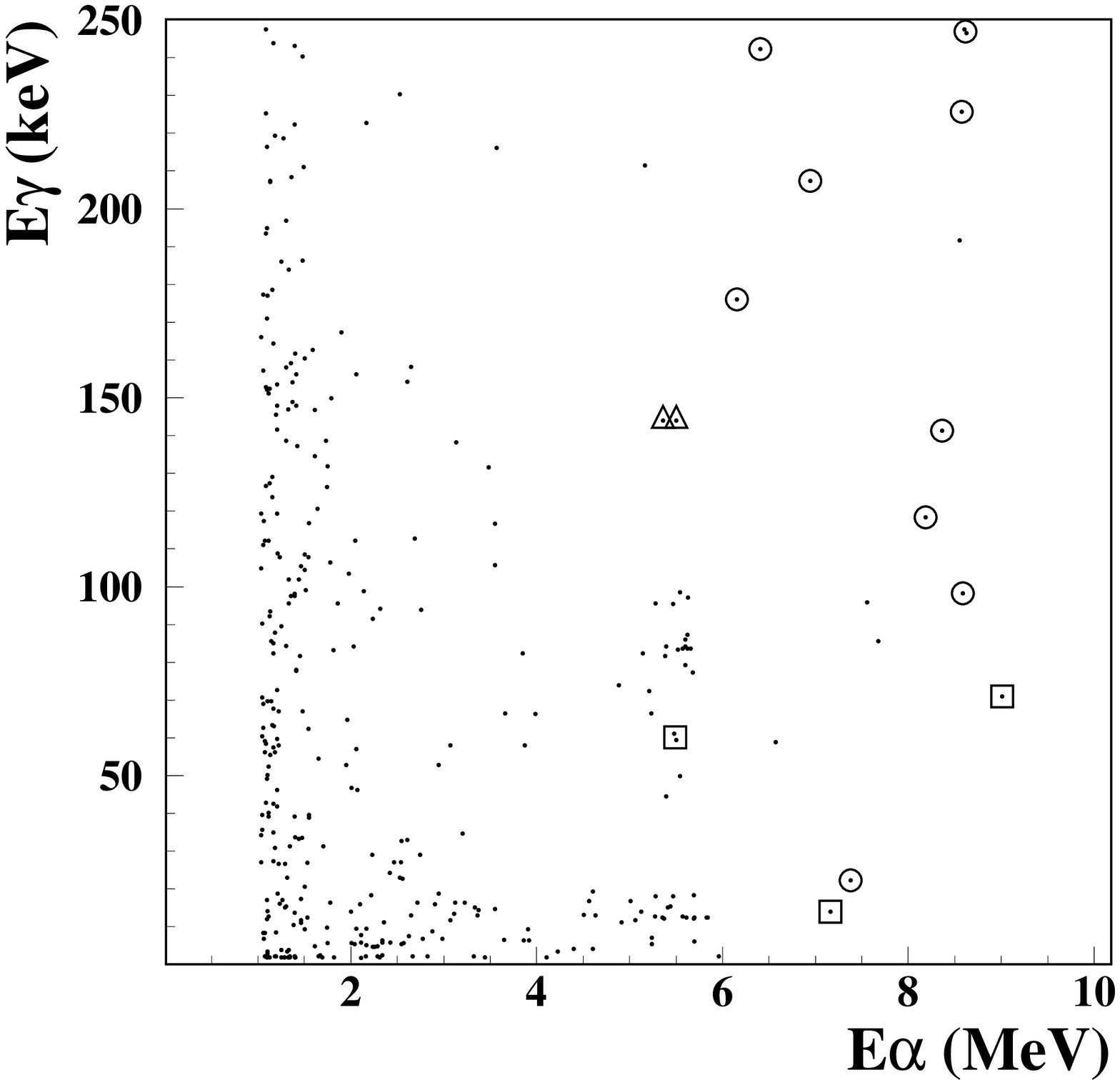}
\end{center}
\caption{Particle-$\gamma$ coincidences from  measurement II:
E$_{Lab}$($^{28}$Si)=125 MeV, with 60 $\mu$g/cm$^2$ C catcher
foil, taken for 96.7 days, starting 3 hours after the end of
irradiation. For further explanations see fig. 5.}
\end{figure}
\begin{figure}
\begin{center}
\leavevmode \epsfysize=8.0cm \epsfbox{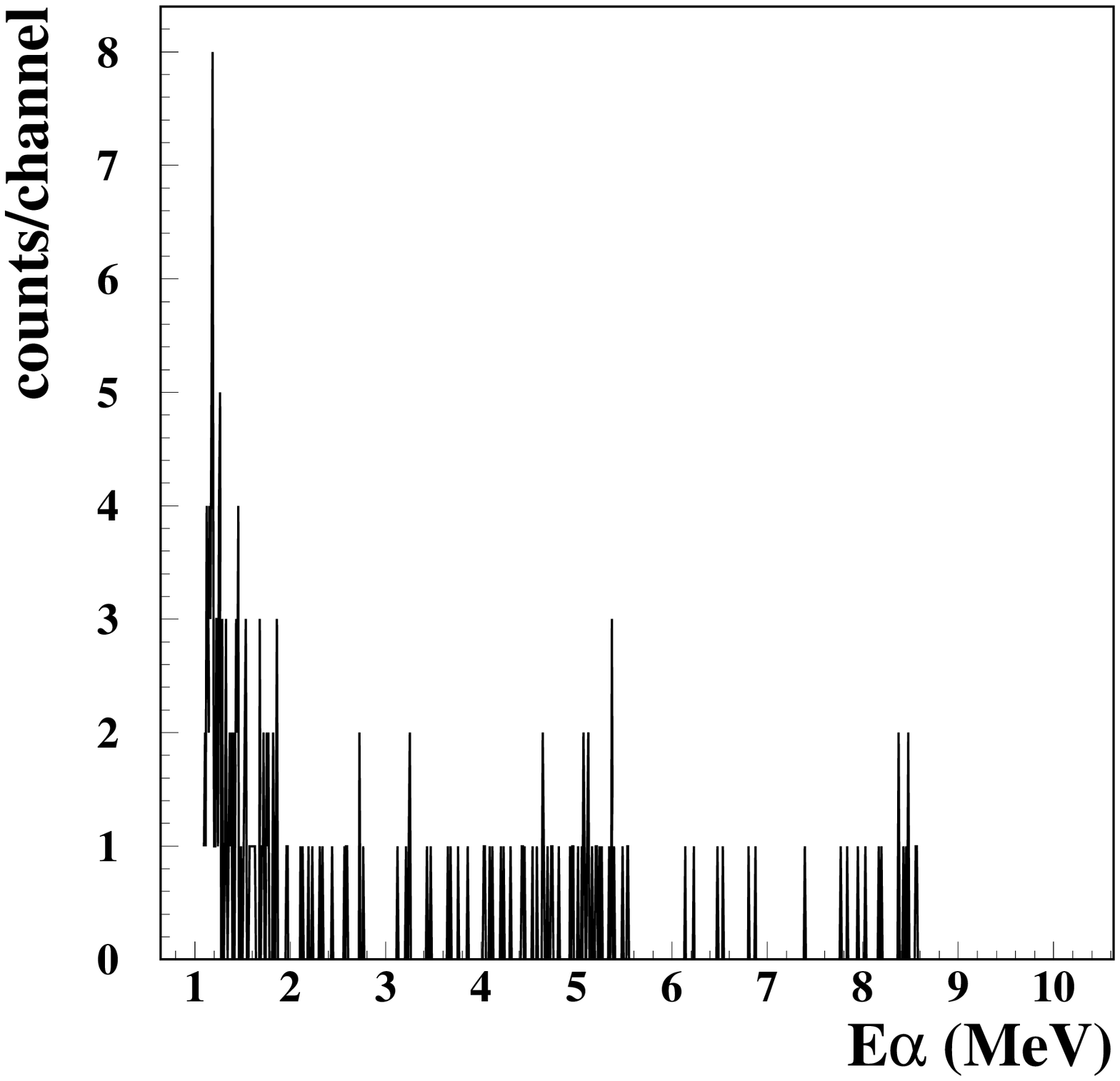}
\end{center}
\caption{Projection of the $\alpha$-$\gamma$ coincidence events
seen in fig.~5 on the $\alpha$-particle axis.}
\end{figure}
\begin{figure}
\begin{center}
\leavevmode \epsfysize=8.0cm \epsfbox{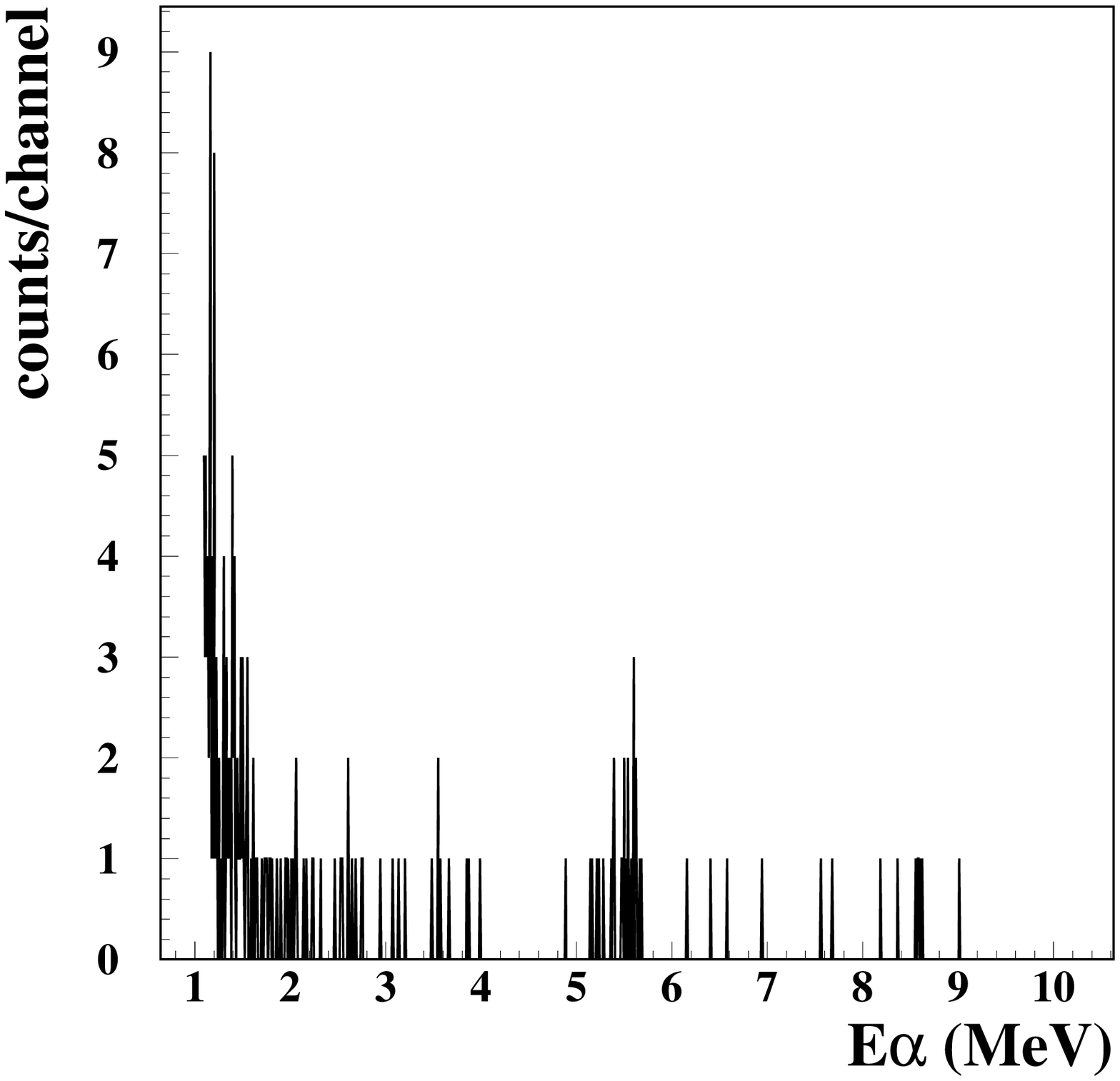}
\end{center}
\caption{Projection of the $\alpha$-$\gamma$ coincidence events
seen in fig.~6 on the $\alpha$-particle axis.}
\end{figure}
\begin{figure}
\begin{center}
\leavevmode \epsfysize=8.0cm \epsfbox{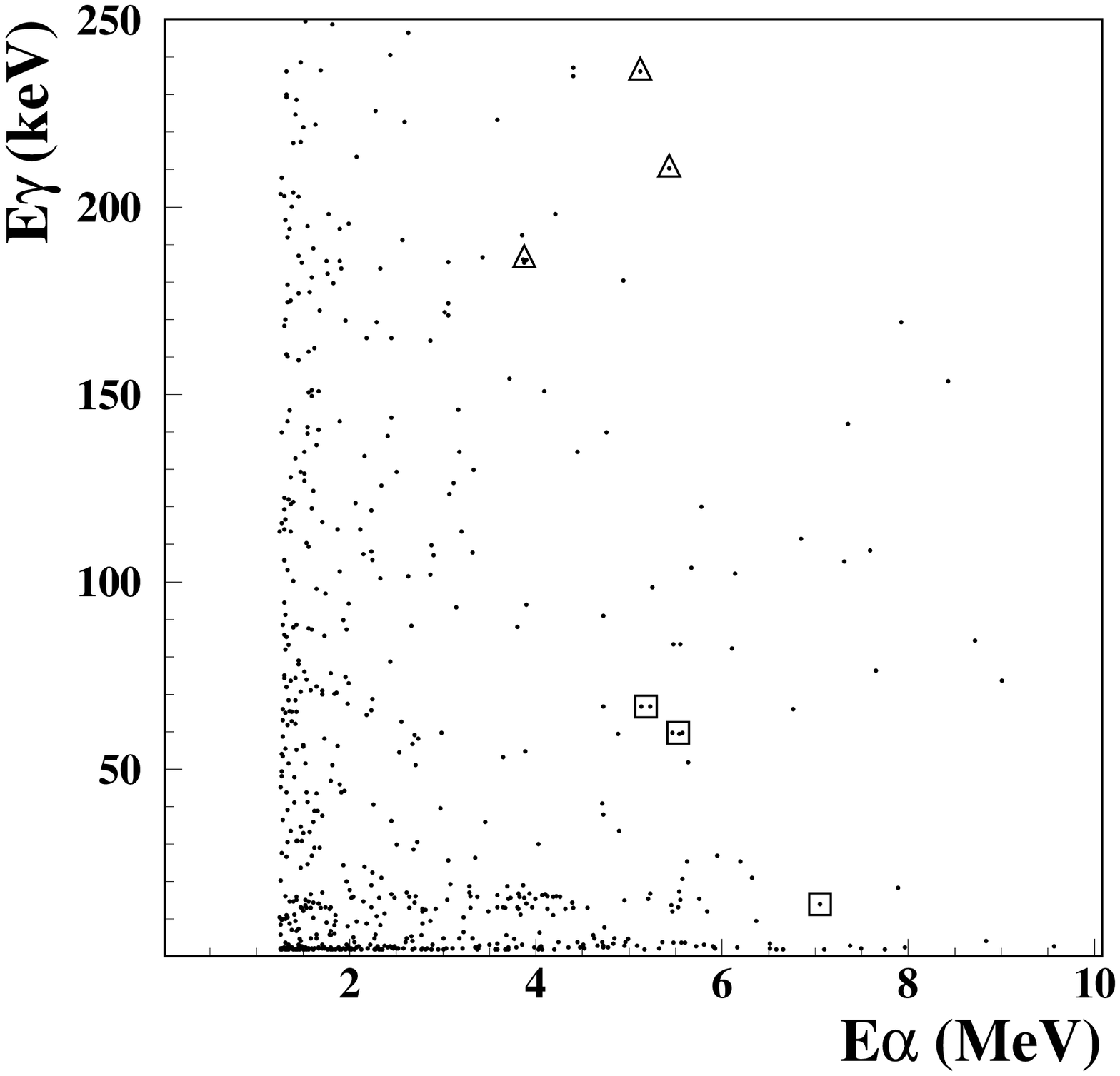}
\end{center}
\caption{Particle-$\gamma$ coincidences from  measurement III:
E$_{Lab}$($^{28}$Si)=135 MeV, with 200 $\mu$g/cm$^2$ C catcher
foil, taken for 235 days, starting 3 hours after the end of
irradiation. For further explanations see fig. 5.}
\end{figure}
\begin{figure}
\begin{center}
\leavevmode \epsfysize=8.0cm \epsfbox{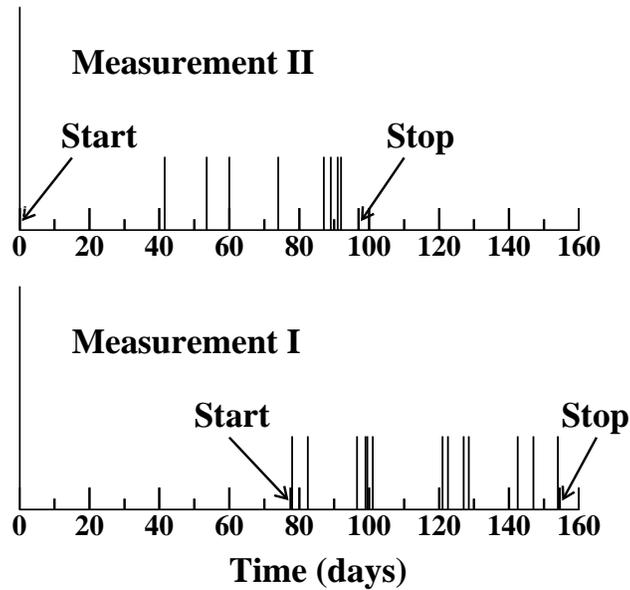}
\end{center}
\caption{Time sequence plots for the $\alpha$-$\gamma$
coincidences with 8 - 9 MeV $\alpha$-particles obtained in
measurements I and II and seen in figs.~5 and 6, respectively.}
\end{figure}
\begin{figure}
\begin{center}
\leavevmode \epsfysize=8.0cm \epsfbox{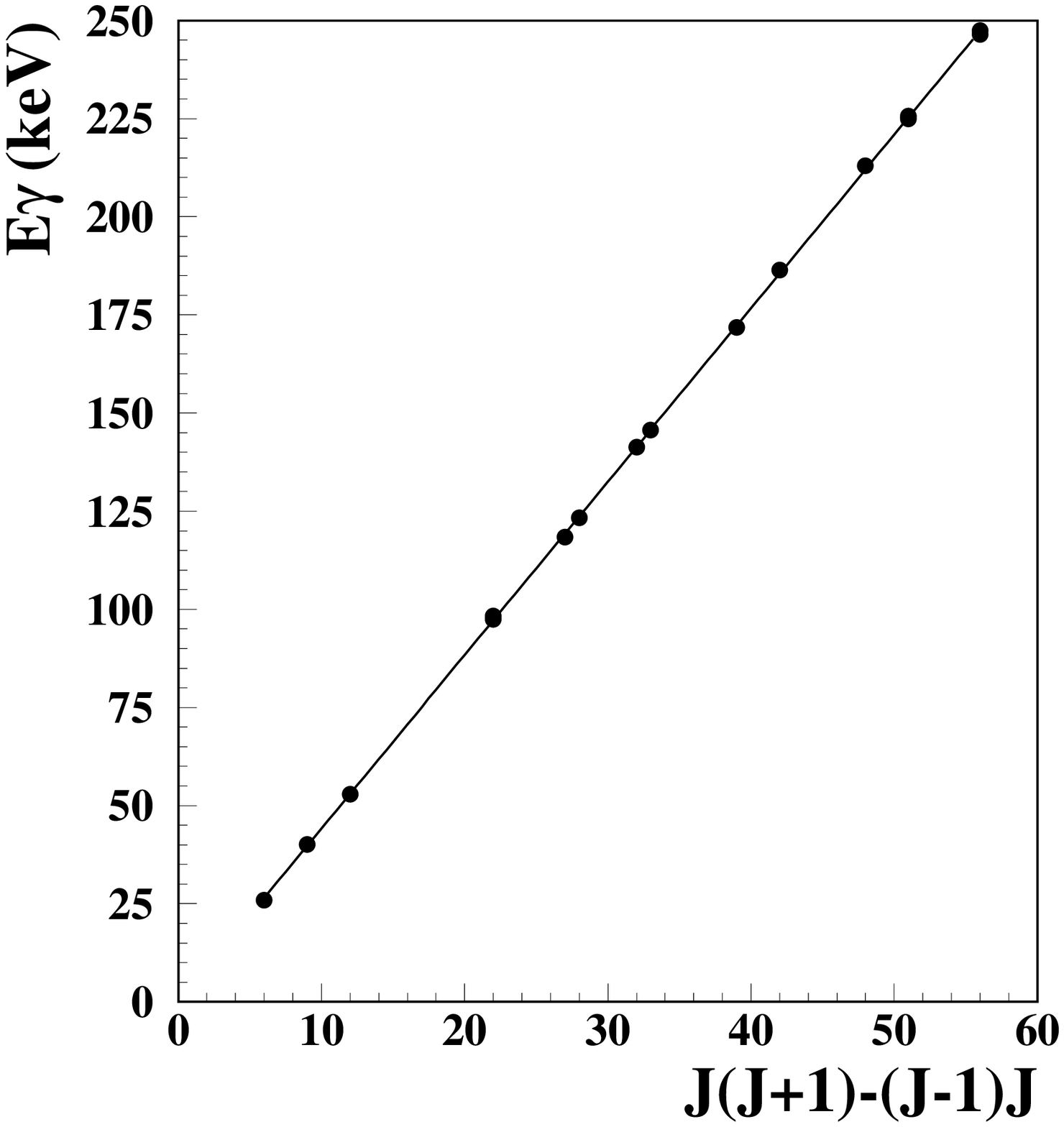}
\end{center}
\caption{E$_\gamma$ versus J(J+1)-(J-1)J for the $\gamma$-rays
seen in coincidence with 7.8--8.6 MeV $\alpha$-particles in
measurements I and II. (Encircled events in figs. 5 and 6, see
table 2(a)).}
\end{figure}


\begin{references}

\bibitem{1} A.~Marinov, S.~Eshhar and D.~Kolb, Phys. Lett. {\bf B191},
36 (1987).
\bibitem{2} A.~Marinov, S.~Eshhar and J. L.~Weil, {\it Proc. Inter. Symp. on
Superheavy Elements, Lubbock, Texas, 1978}, edited by M. H.
K.~Lodhi (Pergamon, New York, 1978), p. 72.
\bibitem{3} A.~Marinov,
C. J.~Batty, A. I.~Kilvington, G. W. A.~Newton, V. J.~Robinson and
J. D.~Hemingway, Nature {\bf 229}, 464 (1971).
\bibitem{4} A.~Marinov, S.~Eshhar and D.~Kolb, Fizika {\bf 19}, Supplement 1,
67 (1987).
\bibitem{5} A.~Marinov, 6th Adriatic Int. Conf. Nucl. Phys.,
 {\it Frontiers of Heavy-Ion
Physics, 1987} Eds. N.~Cindro,
 W.~Greiner and R.~\'{C}aplan,
(World Scientific 1987),
 p. 179.
\bibitem{6} A.~Marinov, S.~Gelberg, and D.~Kolb, Int. Symp. on {\it Exotic
Nuclear States}, Debrecen, Hungary, 1997, Eds. Zs.~Dombr\'{a}di,
Z.~G\'{a}sci and A. Krasznahorkay (akad\'{e}miai Kiad\'{o},
Budapest, APH N.S., Heavy Ion Physics {\bf 7}, 47 (1998).
\bibitem{7} V.~E.~Viola Jr. and G.~T.~Seaborg, J. Inorg. Nucl. Chem. {\bf 28},
741 (1966).
\bibitem{8} A.~Marinov, S.~Gelberg and D.~Kolb, Mod. Phys. Lett.  {\bf A11},
861 (1996).
\bibitem{9} A.~Marinov, S.~Gelberg and D.~Kolb, Mod. Phys. Lett.  {\bf A11},
949 (1996).
\bibitem{10} W.~Satula, S.~\'{C}wiok, W.~Nazarewicz, R.~Wyss and A.~Johnson,
Nucl. Phys. {\bf A529}, 289 (1991).
\bibitem{11} S. J.~Krieger, P.~Bonche, M. S.~Weiss, J.~Meyer,
 H.~Flocard and P.-H.~Heenen, Nucl. Phys. {\bf A542}, 43 (1992).
\bibitem{12} S. G.~Nilsson, G.~Ohl\'{e}n, C.~Gustafson and
P.~M\"{o}ller, Phys. Lett. {\bf 30B}, 437 (1969).
\bibitem{13} J. Fern\'{a}ndez-Neillo, C. H.~Dasso and S.~Landowne,
   Code CCDEF, Comp. Phys. Comm. {\bf 54}, 409 (1985).
\bibitem{14} S. Raman, C. H.~Malarkey, W. T.~Milner, C. W.~Nestor,~Jr.
 and P. H.~Stelson, ADNDT
{\bf 36}, 1 (1987).
\bibitem{15} A.~Marinov, S.~Gelberg and D.~Kolb, {\it Physics of
Unstable Nuclear Beams, S\~{a}o Paulo, Brazil, 1996,} Eds. C.
A.~Bertulani, L.~F.~ Canto and M. S. Hussein, (World Scientific,
1996), p. 181.
\bibitem{16} A.~Marinov, S.~Gelberg and D.~Kolb, $Int. ~School-Seminar ~on
~Heavy ~Ion ~Physics$, Dubna, Russia,1997, Eds. Yu. Ts. Oganessian
R. Kalpakchieva World Scientific, 1998, p. 437.
\bibitem{17} F. S. Stephens et al., Phys. Rev. Lett. {\bf 65}, 301 (1990).
\bibitem{18} S.~$\AA$berg, Nucl. Phys. {\bf A520}, 35c (1990).
\bibitem{19} J.~H\"{o}ller, and S.~$\AA$berg, Z. Phys. {\bf A336}, 363 (1990).
\bibitem{20} W.~Nazarewicz, P.~Olanders, I.~Ragnarsson, J.~Dudek and
G.~A.~Leander, Phys. Rev. Lett. {\bf 52}, 1272 (1984).
\bibitem{21} K.-H.~Schmidt, C.-C.~Sahm, K.~Pielenz and H.-G.~Clerc,
  Z. Phys. {\bf A316}, 19 (1984).
\bibitem{22} G.~Audi, O.~Bersillon, J.~Blachot and A. H.~Wapstra, Nucl. Phys.
{\bf A624}, 1 (1997).
\bibitem{23} R. B.~Firestone, V. S.~Shirley, C. M.~Baglin, S. Y. F.~Chu and
J.~Ziplin, Table of Isotopes, John Wiley and Sons, 1996.
\bibitem{24} W. M.~Howard and P.~M\"{o}ller, ADNDT {\bf 25}, 219 (1980).
\bibitem{25} W.~Nazarewicz, Phys. Lett. {\bf B 305}, 195 (1993).
\bibitem{26} S.~\'{C}wiok, W.~Nazarewicz, J. X.~Saladin, W.~Pl\'{o}ciennik and
A.~Johnson, Phys. Lett. {\bf B 322}, 304 (1994).
\bibitem{27} D.~Kolb and A.~Marinov, Proc. Int. Conf. Nucl. Phys., Florence,
Vol. 1, p. 92 (1983).
\bibitem{28} A.~Marinov, S.~Eshhar, J.~L.~Weil and D.~Kolb, Phys. Rev. Lett.
{\bf 52}, 2209 (1984); {\bf 53}, 1120(E) (1984).
\bibitem{29} A.~Marinov, S.~Gelberg and D.~Kolb, in {\it 6th Int. Conf. on
Nuclei Far from Stability and 9th Int. Conf. on Atomic Masses and
Fundamental Constants}, Bernkastel-Kues, Germany, Inst. Phys.
Conf. Ser. No. 132, p. 437 (1992).
\bibitem{30} K.~Kumar, Superheavy Elements, Adam Hilger, Bristol and New York,
1989.
\bibitem{31} J.~P.~Blocki, H.~Feldmeier and W.~J.~Swiatecki, Nucl. Phys.
{\bf A459}, 145 (1986).

\end{references}
\end{document}